\documentclass[aps, prd, twocolumn, a4paper, 10pt, amsmath, preprintnumbers, nofootinbib, tightenlines, longbibliography, notitlepage,floatfix]{revtex4-1}\pdfoutput=1
\usepackage[english]{babel}
\usepackage{bbm}
\usepackage{graphicx}
\usepackage{hyperref}
\usepackage{wasysym}

\graphicspath{{pics/}}

\newcommand{\abs}[1]{\lvert#1\rvert}

\newcommand{\U}{\Upsilon}
\newcommand{\q}{q}
\newcommand{\qp}{\q_\perp}
\newcommand{\qpi}{\hat{\q}_\perp}
\newcommand{\e}{\epsilon}

\newcommand{\id}[1]{\operatorname{d}\!#1}
\renewcommand{\d}[2]{\frac{\operatorname{d}\!#1}{\operatorname{d}\!#2}}
\newcommand{\dd}[2]{\frac{\operatorname{d}^2\!#1}{\operatorname{d}\!#2^2}}

\newcommand{\ii}{i}
\newcommand{\ee}{e}

\newcommand{\pt}{\tau}

\newcommand{\mt}{\lambda}

\newcommand{\ei}{\hat{\e}}
\newcommand{\mti}{\hat\mt}

\newcommand{\Ui}{\hat{\U} }

\newcommand{\nE}{E}
\newcommand{\nL}{L_z}
\newcommand{\nQ}{Q}

\DeclareMathOperator{\bigO}{O}

\newcommand{\avg}[1]{\langle #1\rangle }

\begin{document}

\title{Resonantly enhanced kicks from equatorial small mass-ratio inspirals}

\author{Maarten \surname{van de Meent}}
\affiliation{Mathematical Sciences, University of Southampton, Southampton, SO17 1BJ, United Kingdom}
\email{M.vandeMeent@soton.ac.uk}

\date{\today}
\begin{abstract}
We calculate the kick generated by an eccentric black hole binary inspiral as it evolves through a resonant orbital configuration where the precession of the system temporarily halts. As a result, the effects of the asymmetric emission of gravitational waves build up coherently over a large number of orbits. Our results are calculated using black hole perturbation theory in the limit where the ratio of the masses of the orbiting objects $\e=m/M$ is small. The resulting kick velocity scales as $\e^{3/2}$, much faster than the $\e^2$ scaling of the kick generated by the final merger. For the most extreme case of a very eccentric ($e\sim 1$) inspiral around a maximally spinning black hole, we find kicks close to $30,000\;\e^{3/2}$~km/s, enough to dislodge an intermediate mass black hole from its host globular cluster. In reality, such extreme inspirals should be very rare. Nonetheless, the astrophysical impact of kicks in less extreme inspirals could be astrophysically significant.
\end{abstract} 

\maketitle
\setlength{\parindent}{0pt} 
\setlength{\parskip}{6pt}

\section{Introduction}
Over the past decade much attention has been drawn to the gravitational recoil or ``kick'' received by merging black hole binaries due to asymmetry in the flux of gravitational waves in the final stages of the merger. Much of this interest is due to the possibility that this kick may be big enough to eject a massive black hole from its host cluster or galaxy \cite{Madau:2004st,Merritt:2004xa}.

In the comparable mass regime, numerical relativity simulations are used to calculate the kick velocities generated by the merger of supermassive black holes \cite{Campanelli:2007ew,*Campanelli:2007cga,Gonzalez:2006md,*Gonzalez:2007hi,Tichy:2007hk,Zlochower:2010sn}. This is supplemented by approximations using post-Newtonian \cite{Blanchet:2005rj,Racine:2008kj} and effective-one-body methods \cite{Damour:2006tr,Schnittman:2007sn}. In the regime where one of the black holes is much heavier than the other, perturbation theory in the small mass ratio $\e=m/M$ may be used to obtain analytical approximations \cite{Fitchett:1984qn,Favata:2004wz,Mino:2008at,Sundararajan:2010sr}.

Most of the work studying kicks in black hole binaries has focussed on the final stage of the merger, starting from the final orbit through the plunge to the ringdown of the merged black hole. This stage is thought to be the most asymmetric and therefore is supposed to produce the biggest kick. Due to relativistic precession any asymmetry in the gravitational radiation emitted earlier during the inspiral phase is supposed to average out over time.

As noted by Hirata \cite{Hirata:2010xn} this is not necessarily the case if there is a resonance between the vertical and azimuthal orbital periods which keeps the system aligned in a certain direction for a prolonged time. He examined the case of strong field inclined circular orbits around a rotating black hole, whose orbital plane precessed an integer number of turns each orbit, i.e. the longitude increase between two ascending nodes is a multiple of $4\pi$. He found that in that case the total recoil velocity from these orbits scaled as $\e^{3/2}$, dominating over the kick from the final merger that scales as $\e^2$  for a large range of ratios.

Strong field orbital resonances themselves have also received their own attention in recent years, as it was found the occurrence of rational ratios of the radial and polar periods of generic (eccentric and inclined) orbits around a rotating black hole can have a significant impact on the evolution of binary inspirals and the resulting gravitational waveform \cite{Flanagan:2010cd,Gair:2011mr,Grossman:2011ps,Flanagan:2012kg,Isoyama:2013yor,Brink:2013nna,Ruangsri:2013hra,Meent:2013}. However, resonances involving the azimuthal period have received very little attention, since the local dynamics cannot directly depend on the value of the azimuthal phase. Nonetheless, as Hirata demonstrated \cite{Hirata:2010xn}, these resonance may still be relevant for quantities that do explicitly depend on the azimuthal angle, like the linear momentum.

This paper will examine the case of orbits for which the radial period is an integer multiple of the azimuthal period, i.e. for which the periapsis shifts by an integer multiple of $2\pi$ each radial period. As with the resonances studied by Hirata, these lead to an enhancement of the gravitational recoil, producing kicks that scale as $\e^{3/2}$. In some sense, this is the simplest type of resonance to examine because it occurs even for equatorial orbits in a Schwarzschild background.

Some work has been done on kicks from eccentric equatorial inspirals in the past \cite{Sopuerta:2006et}. However, like most previous work it focusses on the final merger, and cannot be applied directly to the resonant orbits considered here.

\subsection{Outline of this paper}

The goal of this paper is to calculate the size and direction of the kick generated by a black hole binary inspiral evolving through a $r\phi$-resonance, i.e. a resonance between the radial and azimuthal motion. Section \ref{sec:prelim} discusses the necessary background of the framework describing the evolution of extreme/intermediate mass-ratio inspirals (EMRIs/IMRIs), and introduces some of the used notation. In Sec. \ref{sec:theory} we derive the formulas needed to calculate the total kick incurred as an inspiral evolves through a resonance from the Teukolsky formalism. In addition we discuss the scaling of the kick velocity and the effect it may have on the further evolution of the inspiral. The numerical methods used to solve the Teukolsky equation are briefly outlined in Sec. \ref{sec:nummeth}. Section \ref{sec:results} finally discusses the obtained numerical results.

\subsection{Conventions and Notation}
We employ units such that $c$ (the speed of light), $G$ (Newton's constant), and $M$ (the mass of the central massive black hole) are all unity. Consequently, all quantities appearing in this paper are dimensionless unless specifically noted otherwise. Furthermore the constants of motion $\nE$, $\nL$, and $\nQ$ are normalized to be independent of the invariant particle mass $m$. Metrics have signature  $(- + + +)$. We use the standard Boyer-Lindquist coordinates $(t,r,\theta,\phi)$ for Kerr spacetime.  Without further specification greek indices run over all spacetime coordinates while an index $i$ runs over the subset $(r,\theta,\phi)$. Repeated indices are generally summed over their full range, unless otherwise indicated. The Mino time frequencies $\U_r$ and $\U_\theta$ are considered positive by convention, and $\U_\phi$ can be either positive or negative with $a\U_\phi$ positive for prograde orbits and  $a\U_\phi$ negative for retrograde orbits (where $a$ is the spin parameter of the central massive black hole).

\section{Preliminaries}\label{sec:prelim}
\subsection{Strong field dynamics around a rotating black hole}
In this paper, we study the dynamics of binary systems consisting of two compact objects with masses $M$ and $m$. Moreover, the heavier object of the two ($M$) will be allowed to have a nonzero spin $a=J/M$. We consider the case where the mass ratio $\e\equiv m/M$ is small, so that the corrections to the dynamics can be studied as perturbations in $\e$. At zeroth order, $m$ is a test mass and follows a geodesic in the spacetime geometry generated by $M$, given by the Kerr metric. Higher order corrections appear in the geodesic equation as a force term, the gravitational self-force.

For the analysis of motion in a Kerr background it is convenient to parametrize the worldline of $m$ by the Mino-Carter time \cite{Carter:1968rr,Mino:2003yg}; defined by its relation to proper time $\pt$,
\begin{equation}
\d{\mt}{\pt} =\frac{1}{r(\pt)^2+a^2\cos^2\theta(\pt)},
\end{equation}
where $r(\pt)$ and $\theta(\pt)$ are the position of $m$ in Boyer-Lindquist coordinates. The main advantage of this method is that the geodesic equations with respect to $\mt$ separate into uncoupled equations for $r$ and $\theta$. Moreover, there are some additional analytical and numerical advantages to using $\mt$ as the affine parameter.

Geodesic motion in Kerr background has four constants of motion---the invariant mass $m$, the orbital energy $\nE$, the axial angular momentum $\nL$, and the Carter constant $\nQ$---and is integrable. Consequently, the geodesic equations can be written in terms of action-angle variables $(\q_\mu,J_\mu)$ \cite{Schmidt:2002qk,Hinderer:2008dm},
\begin{subequations}
\begin{align}
\d{q_\mu}{\mt} &= \U_\mu(J_\mu),\\
\d{J_\mu}{\mt} &= 0.
\end{align}
\end{subequations}
The equivalence principle implies that one of these equations is redundant. Moreover, since the Mino time frequencies $(\U_r,\U_\theta,\U_\phi)$ uniquely identify a bound geodesic \cite{Meent:2013}, these equations can be reduced to \cite{Hinderer:2008dm,Meent:2013}
\begin{subequations}
\begin{align}
\d{q_i}{\mt} &= \U_i,\\
\d{\U_i}{\mt} &= 0.
\end{align}
\end{subequations}

The gravitational self-force corrections to the geodesic equation appear as first order corrections to these equations  \cite{Hinderer:2008dm,Meent:2013}
\begin{subequations}
\begin{alignat}{3}
\d{q_i}{\mt} &= \U_i + &\e g_i(\vec\U,q_r,q_\theta) &+\bigO(\e^2),\\
\d{\U_i}{\mt} &= &\e G_i(\vec\U,q_r,q_\theta) &+\bigO(\e^2).
\end{alignat}
\end{subequations}

\subsection{Resonances}
\begin{figure*}[t]
\includegraphics[width=\textwidth]{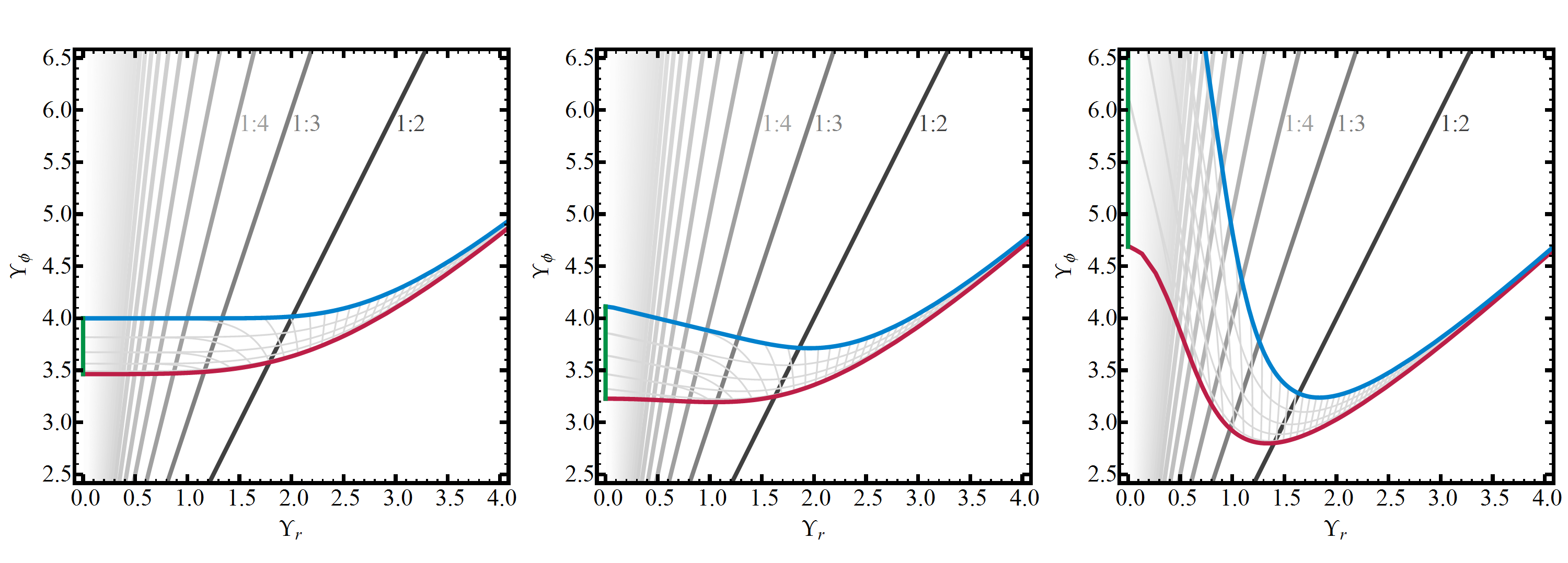}
\caption{The three plots above show the parameter space of eccentric equatorial orbits in terms of the Mino time frequencies for a central black hole with (from left to right) spin $a=0$, $a=0.5$ and $a=0.99$. The grid lines are lines of constant eccentricity or semi-latus rectum. The thick line on the bottom represents the circular orbits, the line on top the parabolic ($e=1$) orbits, and the line on the left (at $\U_r=0$) marks the transition from bound geodesics to plunging geodesics.}\label{fig:rphiplot}
\end{figure*}
Geodesics for which an integer combination of the orbital frequencies $\U$ vanishes,\footnote{Note that the orbital frequencies with respect to Mino time $\vec\U$, coordinate time $\vec\Omega$ and proper time $\vec\omega$ differ from each other by an overall factor. Consequently, the notion of a geodesic being resonant does not depend on the time used to parametrize the geodesic.}
\begin{equation}
\U_\perp\equiv n_\perp\U_r+k_\perp\U_\theta+m_\perp\U_\phi=0,
\end{equation}
are called resonant, where the integers $(n_\perp,k_\perp,m_\perp)$ are chosen to have no common divisors and with $n_\perp\geq0$.\footnote{In the rest of this paper, whenever $X_i$ is a quantity with $i$ running over $(r,\theta,\phi)$, $X_\perp$ will be used to denote the linear combination $n_\perp X_r+k_\perp X_\theta+m_\perp X_\phi$.} Their significance comes from the fact that any quantity oscillating with the frequency $\U_\perp$ will temporarily become constant when an inspiral comes close to a resonant geodesic. Consequently, some oscillating quantities can temporarily show secular growth as the inspiral evolves through the resonance. Generically, this leads to amplification of the relevant effect by a factor $\e^{-1/2}$.

The order of a resonance is defined as $\abs{n_\perp}+\abs{k_\perp}+\abs{m_\perp}$. Higher order resonances generally involve higher order harmonics in the Fourier expansion of quantities, which tend to be exponentially suppressed. Consequently, higher order resonances tend to be less relevant than the lower order ones.

In this paper, we are interested in resonant enhancement of kicks generated by binary inspirals. In \cite{Hirata:2010xn}, Hirata explored this effect for $\theta\phi$-resonances (i.e. resonances with $n_\perp=0$) in quasi-circular non-equatorial orbits around a rotating black hole, noting that a similar effect should occur for  $r\phi$-resonances (i.e. $k_\perp=0$). In this paper we examine resonant enhancement of kicks by $r\phi$-resonances in eccentric equatorial orbits.

Figure \ref{fig:rphiplot} shows the parameter space of eccentric equatorial geodesics in terms of the Mino time frequencies $\U_r$ and $\U_\phi$ for various values of the spin $a$ of the central black hole. We want to stress a number of features of the $r\phi$-resonances based on these plots. First, inspirals start in the top right region of each of the plots and evolve to the left, with the transition to plunge happening at the $\U_r=0$ axis. In particular, a quasi-circular inspiral simply follows the bottom line in each diagram. Consequently, all inspirals encounter all $r\phi$-resonances. This in contrast to the  $\theta\phi$-resonances which occur only for inspirals with specific initial conditions. Second, the  $r\phi$-resonances happen well before the final plunge of the binary system. Consequently, the kick produced by evolution through an $r\phi$-resonance is distinct from the kick produced by the final merger. This in contrast to the $\theta\phi$-resonances, which may happen right before the plunge, meaning that in numerical simulations their effect could be confused with the kick from the final merger. Finally, as seen in the right most diagram, for an extremal black hole the $\phi$ frequency for very eccentric orbits blows up near the final plunge. This is a result of the geodesic becoming nearly null as it whirls around the light-ring.  However, the most relevant lower order resonances occur well away from this singular behaviour.

\subsection{Eccentric orbits}
For orbits in the strong field regime we define eccentricity in analogy with the Newtonian terminology. The Boyer-Lindquist radial coordinate of a bound geodesic in Kerr spacetime has two turning points $r_\textrm{min}$ and $r_\textrm{max}$. The eccentricity $e$ and the semi-latus rectum $p$ are defined by imposing the familiar Newtonian relations with the radial turning points,
\begin{subequations}
\begin{align}
 r_\textrm{max} &= \frac{p}{1-e},\\
 r_\textrm{min} &= \frac{p}{1+e}.
\end{align}
\end{subequations}

\section{Theory}\label{sec:theory}
\subsection{Symmetry breaking}
The equations of motion for a particle moving in Kerr spacetime obey the same symmetry rules as the background metric, i.e. they are invariant under time translations, axial-rotations, and mirroring in the equatorial plane (up-down symmetry). Generic (non-resonant) bound solutions to the geodetic equations obey the same symmetries in the sense that a rotation or up-down exchange is equivalent to an appropriate shift of the affine parameter. Consequently, the long-time average flux of gravitational radiation to infinity will also be invariant under axial rotation and up-down symmetry. Since linear momentum is not invariant under these operations, the average linear momentum carried by the flux of gravitational waves must be zero.

To produce a nonzero net flux of linear momentum in the equatorial plane, a geodesic must break the rotational invariance of the background. This happens for geodesics with a resonance between the radial and azimuthal frequencies $\U_r$ and $\U_\phi$, i.e. orbits for which the periapsis shift is a multiple of $2\pi$. Note that the rotational symmetry needs to be broken completely, since a residual discrete symmetry would still imply a zero net flux of linear momentum in the equatorial plane. This implies that $m_\perp=-1$. This option will be examined in this paper. 

Similarly, the production of a net flux of the linear momentum in the axial direction requires the breaking of up-down symmetry. This can happen for some geodesics with a resonance between the radial and polar motion. Whether or not up-down symmetry is broken will depend not only on the ratio of the frequencies, but also on the relative phase of the radial and polar motion. For example, a $r\theta$-resonant orbit with $n_\perp$=2 and $k_\perp=-1$ will break up-down symmetry when the polar motion reaches its extreme at periapsis, while up-down symmetry remains intact if the geodesic passes through the equator at periapsis. As discussed in \cite{Meent:2013}, there may exist the possibility that the radiation reaction driven evolution gets ``stuck'' on a $r\theta$-resonance. If this were to happen in conjunction with the breaking of up-down symmetry, this could lead to a dramatic buildup of linear momentum in the axial direction. This possibility will be examined in a future paper.

\subsection{Linear momentum flux to infinity}\label{sec:linmom}
All information about gravitational radiation travelling out to infinity is encoded in the Penrose scalar $\psi_4$. Teukolsky showed \cite{Teukolsky:1973ha} that the partial differential equation for perturbations of   $\psi_4$ separates if one imposes a Fourier and multipolar decomposition,
\begin{equation}\label{eq:TeukDecomp}
\psi_4 = \frac{1}{(r-\ii a \cos\theta)^4}\sum_{lm\omega} R_{lm\omega}(r)S_{lm\omega}(\Omega)\exp(\ii \omega t),
\end{equation} 
where the $S_{lm\omega}(\Omega)$ are eigenfunctions of the angular equation, called spin-weighted spheroidal harmonics, the $ R_{lm\omega}(r)$ are solutions of the (radial) Teukolsky equation, and we have anticipated that the sources terms that we will be using (generated by bound geodesics) have a discrete $\omega$ spectrum.

For $r$ outside the source region, the source term in the Teukolsky equation vanishes and $ R_{lm\omega}(r)$ is proportional to a solution of the homogeneous Teukolsky equation, i.e.
\begin{equation}\label{eq:TeukOut}
 R_{lm\omega}(r) = Z^{\mathrm{out}}_{lm\omega} R^{\mathrm{out}}_{lm\omega}(r),
\end{equation} 
where $ R^{\mathrm{out}}_{lm\omega}(r)$ is the homogeneous solution of the Teukolsky equation that has no radiation coming in from (past null) infinity, and $Z^{\mathrm{out}}_{lm\omega}$ is a constant of proportionality.

The total flux of 4-momentum flowing out to infinity can be calculated by projecting the energy flux at infinity onto a Cartesian frame vector $n_\mu=(-1, \sin\theta\cos\phi,\sin\theta\sin\phi,\cos\theta)$ with $\mu$ running over $(t,x,y,z)$,
\begin{equation}
\d{P_\mu}{t} = \int_{S^2} n_\mu  \lim_{r\to\infty} r^2 \avg{\abs{\int_{-\infty}^t \psi_4\id{t'}}^2 } \id\Omega,
\end{equation} 
where the angular brackets $\avg{}$ indicate averaging of a timescale long (possibly infinite) with respect to the period of the gravitational waves. Since the perturbation to curvature scalar is of order $\e$, we find that the change of the linear momentum is of order $\e^2$.

Using \eqref{eq:TeukDecomp} and \eqref{eq:TeukOut}, the angular distribution of gravitational radiation at infinity can be expressed as,
\begin{multline}
 \lim_{r\to\infty} r^2 \avg{\abs{\int_{-\infty}^t \psi_4\id{t'}}^2 } = \\
 \sum_{\substack{lm\omega\\l'm'\omega'}}
  \frac{Z^{\mathrm{out}}_{lm\omega}\bar{Z}^{\mathrm{out}}_{l'm'\omega'}}{\omega\omega'}
  S_{lm\omega}(\Omega)\bar{S}_{l'm'\omega'}(\Omega)\delta_{\omega\omega'},
\end{multline}
where we used that the time integral just multiplies each component of $\psi_4$ in  \eqref{eq:TeukDecomp} by $\ii/\omega$, and the average results in a Kronecker delta $\delta_{\omega\omega'}$ forcing the $\omega$'s to match. If the source is a non-resonant orbit this immediately implies that $m=m'$ as well, because
\begin{equation}
\omega=\Omega_\phi m +\Omega_r n +\Omega k = \Omega_\phi m' +\Omega_r n' +\Omega k'=\omega' 
\end{equation} 
can only be satisfied if $m=m'$, $n=n'$, and $k=k'$. For a  resonant orbit specified by the integers $ (m_\perp,n_\perp,k_\perp)$, this condition reduces to just
\begin{equation}
(m-m',n-n',k-k')=N (m_\perp,n_\perp,k_\perp),
\end{equation} 
for some integer $N$.

The flux of linear momentum in the orbital plane is given by
\begin{align}
\d{P_{xy}}{t}&\equiv\d{P_x}{t}+\ii\d{P_y}{t} =
\\&\hspace{-1.0em}=- \sqrt{\frac{8\pi}{3}} \int_{S^2}\!\!\!  Y_{11}  \lim_{r\to\infty} r^2 \avg{\abs{\int_{-\infty}^t \psi_4}^2 \id{t}}\id\Omega\\ 
&\hspace{-1.0em}=- \sqrt{\frac{8\pi}{3}}\! \sum_{\substack{lm\\l'm'}\omega}\!\!
  \frac{Z^{\mathrm{out}}_{lm\omega}\bar{Z}^{\mathrm{out}}_{l'm'\omega}}{\omega^2}
 \!\!\int_{S^2}\!\!\! Y_{11} S_{lm\omega}\bar{S}_{l'm'\omega} \id{\Omega}. \label{eq:Pxyflux1}
\end{align} 
By decomposing the spin-weighted spheroidal harmonic in spin-weighted spherical harmonics \cite{Hughes:1999bq},
\begin{equation}
 S_{lm\omega} = \sum_j b^{j}_{lm\omega} {^{-2}Y_{jm}},
\end{equation} 
the last integral in \eqref{eq:Pxyflux1} can be evaluated explicitly in terms of Wigner $3j$ symbols.
\begin{align}
 \int_{S^2}\!\!\! Y_{11} S_{lm\omega}\bar{S}_{l'm'\omega} \id{\Omega}&=\\
 &\hspace{-7.6em}= \sum_{j,j'} b^{j}_{lm\omega}\bar{b}^{j'}_{l'm'\omega}\int_{S^2}\!\!\! Y_{11}{^{-2}Y_{jm}}{^{-2}\bar{Y}_{j' m'}} \id\Omega\\
 \begin{split}
 &\hspace{-7.6em}=(-1)^{m'} \sum_{j,j'} b^{j}_{lm\omega}\bar{b}^{j'}_{l'm'\omega}\sqrt{\frac{3(2j+1)(2j'+1)}{4\pi}}\\
 &\times\begin{pmatrix}
1 & j & j'\\1 & m & -m'
\end{pmatrix}
\begin{pmatrix}
1 &j &j'\\0 & 2 & -2
\end{pmatrix}. 
\end{split} 
\end{align} 
The Wigner $3j$ symbols in the last line vanish unless $m-m' = 1$ and $j-1\leq j' \leq j+1$. A direct consequence of the first condition is that $\d{P_x}{t}+\ii\d{P_y}{t}$ can only be zero for resonances with $m_\perp=-1$.

Similarly the axial component of the linear momentum flux is given by
\begin{align}
\d{P_z}{t}&\hspace{0em}= \sqrt{\frac{4\pi}{3}} \int_{S^2}\!\!\! Y_{10}  \lim_{r\to\infty} r^2 \avg{\abs{\int\id{t} \psi_4}^2 }\id\Omega \\ 
&\hspace{0em}=\sqrt{\frac{4\pi}{3}} \sum_{\substack{lm\\l'}\omega}
  \frac{{Z^{\mathrm{out}}_{lm\omega}\bar{Z}^{\mathrm{out}}_{l'm\omega}}}{\omega^2}
 \int_{S^2}\!\!\! Y_{10} S_{lm\omega}\bar{S}_{l'm\omega} \id{\Omega},
\end{align} 
where we have used that explicit calculation of the integral in the last line,
\begin{align}
 \int_{S^2}\!\!\! Y_{10} S_{lm\omega}\bar{S}_{l'm'\omega} \id{\Omega}&=\\
 &\hspace{-7.6em}= \sum_{j,j'} b^{j}_{lm\omega}\bar{b}^{j'}_{l'm'\omega}\int_{S^2}\!\!\! Y_{10}{^{-2}Y_{jm}}{^{-2}\bar{Y}_{j' m'}} \id\Omega\\
 \begin{split}
 &\hspace{-6.9em}=(-1)^{m'} \sum_{j,j'} b^{j}_{lm\omega}\bar{b}^{j'}_{l'm'\omega}\sqrt{\frac{3(2j+1)(2j'+1)}{4\pi}}\\
 &\times\begin{pmatrix}
1 & j & j'\\0 & m & -m'
\end{pmatrix}
\begin{pmatrix}
1 &j &j'\\0 & 2 & -2
\end{pmatrix},
\end{split} 
\end{align}
implies that $m=m'$.
\subsection{Total kick}\label{sec:totkick}
As long as the evolution is slow enough, the total linear momentum expelled by a binary inspiral as it evolves through a resonance can be calculated from the dependence of the linear momentum flux from a resonant geodesic on the value of the resonant phase $\qp = n_\perp \q_r + k_\perp \q_\theta+m_\perp \q_\phi$. This is a general feature of evolution through resonance stressed in \cite{Meent:2013}. In Appendix \ref{app:Xkick}, we derive a general formula for the total kick to quantities that do not directly appear in the equations of motion, such as the linear momentum.

In this paper we focus on $r\phi$-resonances that break rotational invariance. In that case, recall that $k_\perp = 0$ and $m_\perp=-1$, so that $\qp = n_\perp \q_r -\q_\phi$. Moreover, since $\qp$ is the phase that is constant on a resonant orbit, we can evaluate it at any point of the orbit we choose. In particular, at periapsis $\q_r=0$, so, $\qp = -\q_{\phi,0}$, where $\q_{\phi,0}$ is the value of $\q_\phi$ at periapsis. Finally, from the analytic solutions to the geodesic equations described in \cite{Fujita:2009bp} it follows that  $\q_{\phi,0}$ is equal to $\phi_0$; the value of the axial Boyer-Lindquist coordinate $\phi$ at periapsis.

Finding the dependence of $\d{P_{xy}}{t}$ on $\phi_0$ is straightforward. If $(\d{P_{xy}}{t})_0$ is the total linear momentum flux for a resonant geodesic with $\phi_0=0$, then by rotational invariance of the Kerr background the linear momentum flux for a geodesic with a different value of $\phi_0$ is $\exp(\ii\phi_0)(\d{P_{xy}}{t})_0$. Consequently,
\begin{equation}
\d{P_{xy}}{\mt}(\qp) = \exp(-\ii\qp)(\d{P_{xy}}{t})_0\Gamma,
\end{equation}
where $\Gamma= \avg{\d{t}{\mt}}$.
Consequently, if we write,
\begin{equation}
\d{P_{xy}}{\mt}(\qp) = \e^2 \sum_{N\neq 0} F_N \exp(\ii N\qp),
\end{equation}
we find that all $F_N$ vanish, except
\begin{equation}
F_{-1} = \e^{-2}(\d{P_{xy}}{t})_0\Gamma.
\end{equation}
Plugging this into Eq. \eqref{eq:totalDX2} we find the total linear momentum emitted by the system as the inspiral evolves through a $r\phi$-resonance,
\begin{equation}
\Delta{P_{xy}} = \e^{3/2}\sqrt{2\pi} \frac{F_{-1}}{\abs{\avg{G_\perp}}^{1/2}}\ee^{\ii(\phi_0-\pi/4)}.
\end{equation}
Here, $F_{-1}$ can be computed through the methods of the previous section while $\avg{G_\perp}$ can be obtained from the infinite time averages of the total fluxes of the energy and (axial) angular momentum to infinity and down the black hole horizon using the method of appendix A of \cite{Meent:2013}.

Since Kerr spacetime is asymptotically flat, the ADM style total linear momentum is conserved. Consequently, the velocity of the center of mass of the binary system has to change in the opposite direction to the emitted linear momentum. The total kick velocity $\vec{V}^\textrm{kick}$ that the binary system recieves as it evolves through a $r\phi$-resonance is thus given by
\begin{equation}\label{eq:kickvel}
\begin{split}
V^\textrm{kick}_x+\ii V^\textrm{kick}_y &=\\
&\hspace{-1.5em}c \e^{3/2}\sqrt{2\pi} \frac{F_{-1}}{\abs{{\avg{G_\perp}}}^{1/2}}\ee^{\ii(\phi_0+3\pi/4)}+\bigO(\e^2),
\end{split}
\end{equation}
where we have restored all dimensional quantities. At leading order the kick velocity is independent of the total mass of the binary system; it only depends on the mass ratio. More importantly, the magnitude of the kick scales with $\e^{3/2}$, which is a stronger scaling than the $\e^{2}$ scaling of kick effects from plunge, merger, and ringdown. Consequently, the resonant kick is expected to dominate over the kick from the final merger for small mass ratios.

It is easy to see that this scaling cannot persist all the way to the comparable mass regime, since for exactly equal masses symmetry is restored and there can be no net kick effect in the orbital plane. The perturbation expansion must therefore breakdown before the comparable mass regime.

Moreover, at some mass-ratio the orbital time, resonance, and evolution time scales all become similar and the assumption that the evolution is slow compared to the orbital time scale fails. Appendix \ref{app:Xkick} derives an upper bound $\e_\textrm{crit}$ on the mass ratio where this assumption becomes false,
\begin{equation}\label{eq:epscrit}
\e \ll \e_{\mathrm{crit}}\equiv\frac{\U_r^2}{\pi \abs{\avg{G_\perp}}}.
\end{equation}

\subsection{Effect on orbital evolution}
Conventional wisdom is that resonances involving the azimuthal frequency cannot affect the evolution of the binary system because the equations of motion do not depend on $\q_\phi$. In particular, $r\phi$-resonances should not cause jumps in the constants of motion $E$, $L$, and $Q$. We will argue here that while this is true at leading order, a kick to the velocity of the center of mass of the system also implies a (small, higher order) correction to the orbital constants of motion.

The easiest way to understand this is in terms of the energy balance of the system. Ordinary flux balance arguments assume that on average the total energy carried away by gravitational waves is equal to the change of the ``internal'' energy of the binary system, which can be split in the change of mass of the central object (which can be calculated from the gravitational wave flux down the horizon) and the orbital energy (expressed by the constant of motion $E$). However, if a kick changes the velocity of the center of mass, then there is also energy going into the kinetic energy of the system. Correspondingly, the orbital energy $E$ of the system after the kick will be smaller than what one would expect if the $r\phi$-resonance causing the kick were ignored.

To get a handle on the size of this effect suppose that the kick velocity $v_{\mathrm{kick}}$ is small compared to the speed of light, so that a distant observer can treat the energy balance of the system non-relativistically. In that case the total energy of the system before and after the kick can be written,
\begin{align}
E^\mathrm{tot}_\mathrm{before} &= M_\mathrm{before} + E^\mathrm{orb}_\mathrm{before} \\
E^\mathrm{tot}_\mathrm{after} &= M_\mathrm{after} + E^\mathrm{orb}_\mathrm{after} + E^\mathrm{GW}_\mathrm{\infty} + \frac{1}{2}M_\mathrm{after}v_{\mathrm{kick}}^2.
\end{align} 
Consequently, the change in orbital energy is
\begin{align}
\Delta E^\mathrm{orb} &= E^\mathrm{orb}_\mathrm{after}- E^\mathrm{orb}_\mathrm{before}\\
&= M_\mathrm{before} -M_\mathrm{after} - E^\mathrm{GW}_\mathrm{\infty} - \frac{1}{2}M_\mathrm{after}v_{\mathrm{kick}}^2\\
&= -E^\mathrm{GW}_\mathrm{hor} - E^\mathrm{GW}_\mathrm{\infty} - \frac{1}{2}M_\mathrm{after}v_{\mathrm{kick}}^2.
\end{align} 
As argued elsewhere in this paper the total gravitational wave flux emitted as a system evolves through a $r\phi$-resonance is of order $\e^{3/2}$. The total kick velocity  $v_{\mathrm{kick}}$ is of similar order, so the correction to the energy due to the kick is of order $\e^3$. The correction to the orbital frequencies is therefore order $\e^2$. In the remainder of the evolution of the inspiral from the resonance to the final plunge into the central object, this accumulates to a correction to the orbital phases of order $\e$. Consequently, it is probably safe to ignore this correction for the analysis of gravitational waveforms from IMRIs and EMRIs. Nonetheless, the effect is of similar order as (non-secular) second order self-force corrections. A full consistent treatment of the second order gravitational self-force is thus expected to include this type of correction.

\section{Numerical Methods}\label{sec:nummeth}
To numerically obtain $\psi_4$ we have implemented the methods described in \cite{Fujita:2004rb,Fujita:2009bp,Fujita:2009us,ST:lrr-2003-6,Throwethesis} in \emph{Mathematica}. The process is as follows.

We first pick a spin $a$ for the central black hole, an eccentricity $e$ for the orbit and the resonant ratio $R=\U_\phi/\U_r$ to fix a resonant equatorial orbit. We then numerically invert the analytical expressions for Mino time frequencies found in \cite{Fujita:2009bp} to obtain the semilatus rectum $p$ of that orbit. The parameters $(a,e,p)$ are then used to construct the analytical solution geodetic orbit with those parameters  \cite{Fujita:2009bp}, which we expand as a Fourier series in Mino time.

This orbit is then used to construct a source for the Teukolsky equation, which we solve mode by mode using the method of variation of parameters. Homogeneous solutions to the Teukolsky equation representing outgoing waves at infinity and down going waves at the black hole horizon are constructed as series of hypergeometric functions as described in \cite{Fujita:2004rb}. To compute the inhomogeneous solution an integral over the region containing the source is required. Evaluating the series of hypergeometric functions at many points in that region is computationally very prohibitive. So, instead we use the series of hypergeometric functions to obtain very precise values of the homogeneous solutions and their first derivatives at one point in the source region. We then use standard Taylor series methods to extend this solution to the entire source region as described in \cite{Fujita:2009us} and use that to perform the integral.

This gives us the asymptotic amplitudes of the inhomogeneous solutions of the Teukolsky equation, which are used as input for the formulas in Sec. \ref{sec:linmom} the flux of linear momentum to infinity, and the total change of energy and angular momentum. The last two can be used to obtain the change of the resonant frequency $\d{\U_\perp}{\mt}$.

This gives us all the ingredients necessary to compute the total kick to the linear momentum incurred as the system evolves through the specified $r\phi$-resonance as described in Sec. \ref{sec:totkick}.

\section{Results}\label{sec:results}
\begin{figure}
\includegraphics[width=\columnwidth]{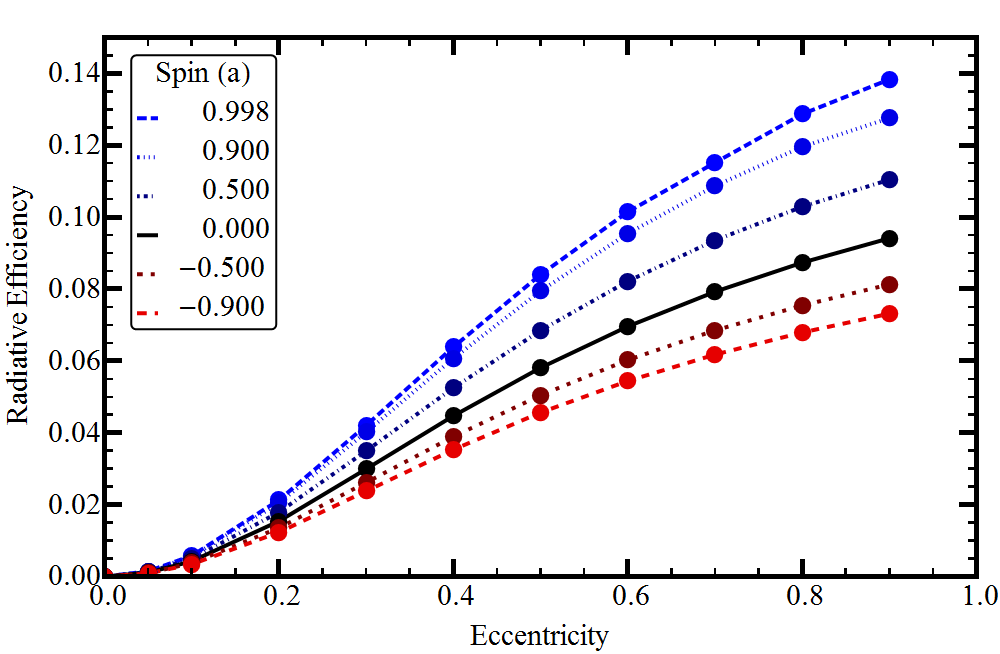}
\caption{The radiative efficiency $\abs{\d{\vec{P}}{t}}/\d{\nE}{t}$ of various orbits in 1:2 $r\phi$-resonance.}\label{fig:radeff}
\end{figure}
We first examine the effect of $r\phi$-resonances with a 1 to 2 frequency ratio. This is the first $r\phi$-resonance any inspiral will encounter. And as noted before, it also is expected to be the strongest. We examine a range of equatorial orbits with eccentricity varying between $0$ and $0.9$, around central objects with spin $a/M$ varying between $-0.9$ (i.e. anti-aligned with the orbital angular momentum) and 0.998. Figure \ref{fig:radeff} shows the efficiency of the gravitational wave rocket effect for each orbit, i.e. the ratio of $\abs{\d{P_{xy}}{t}}$ by the total energy flux to infinity. As expected, it is zero for circular orbits. For small eccentricity it grows approximately as $e^2$. It maxes out at approximately $15\%$ for prograde (nearly) parabolic orbits around a maximally spinning central black hole. This is much more efficient than the $\theta\phi$-resonances studied by Hirata \cite{Hirata:2010xn}, which had an efficiency of about $0.6\%$. 

\begin{figure}
\includegraphics[width=\columnwidth]{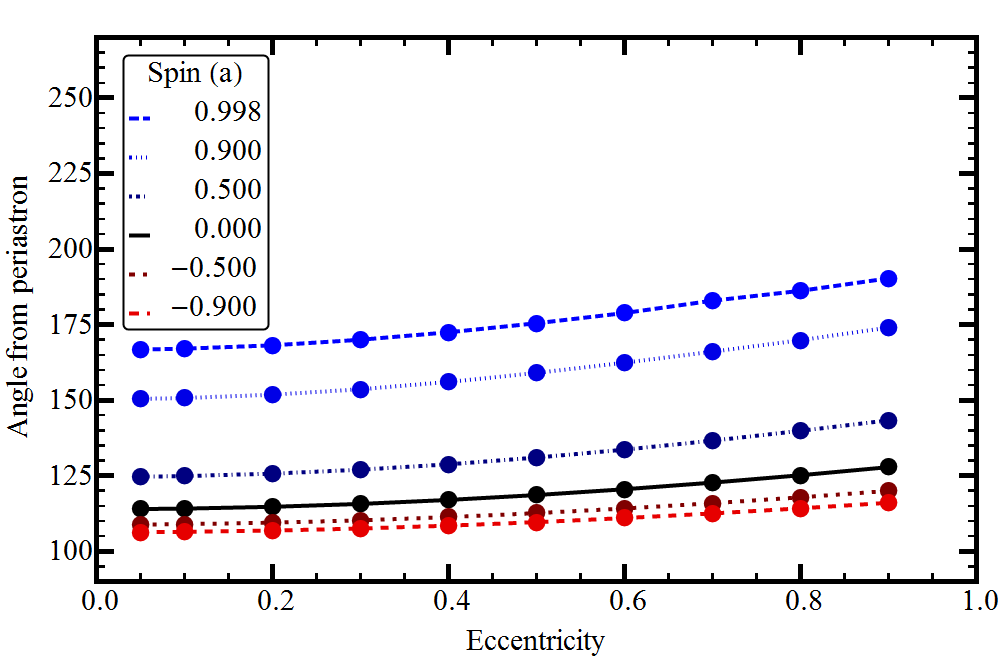}
\caption{The net direction of the radiative inbalance of the orbits in Fig. \ref{fig:radeff}, measured as the angle from periapsis.}\label{fig:inbaldir}
\end{figure}

Figure \ref{fig:inbaldir} shows the direction of the radiative imbalance, i.e. the argument of $\d{P_{xy}}{t}$. The direction is expressed as the angle from periapsis in the orbital direction. As a rough expectation, one would assume that the main imbalance comes from the difference between the radiation emitted at periapsis and apoapsis. Since gravitational radiation is focussed in the direction of motion, this would imply that the direction of imbalance would be at approximately 90 degrees from periapsis. The numerical results show that this is almost true for the retrograde orbits, but for the prograde orbits around highly spinning black holes the direction increases to 180 degrees. The main cause for this is deflection of the gravitational waves in the curved background.
\begin{figure}[b]
\includegraphics[width=\columnwidth]{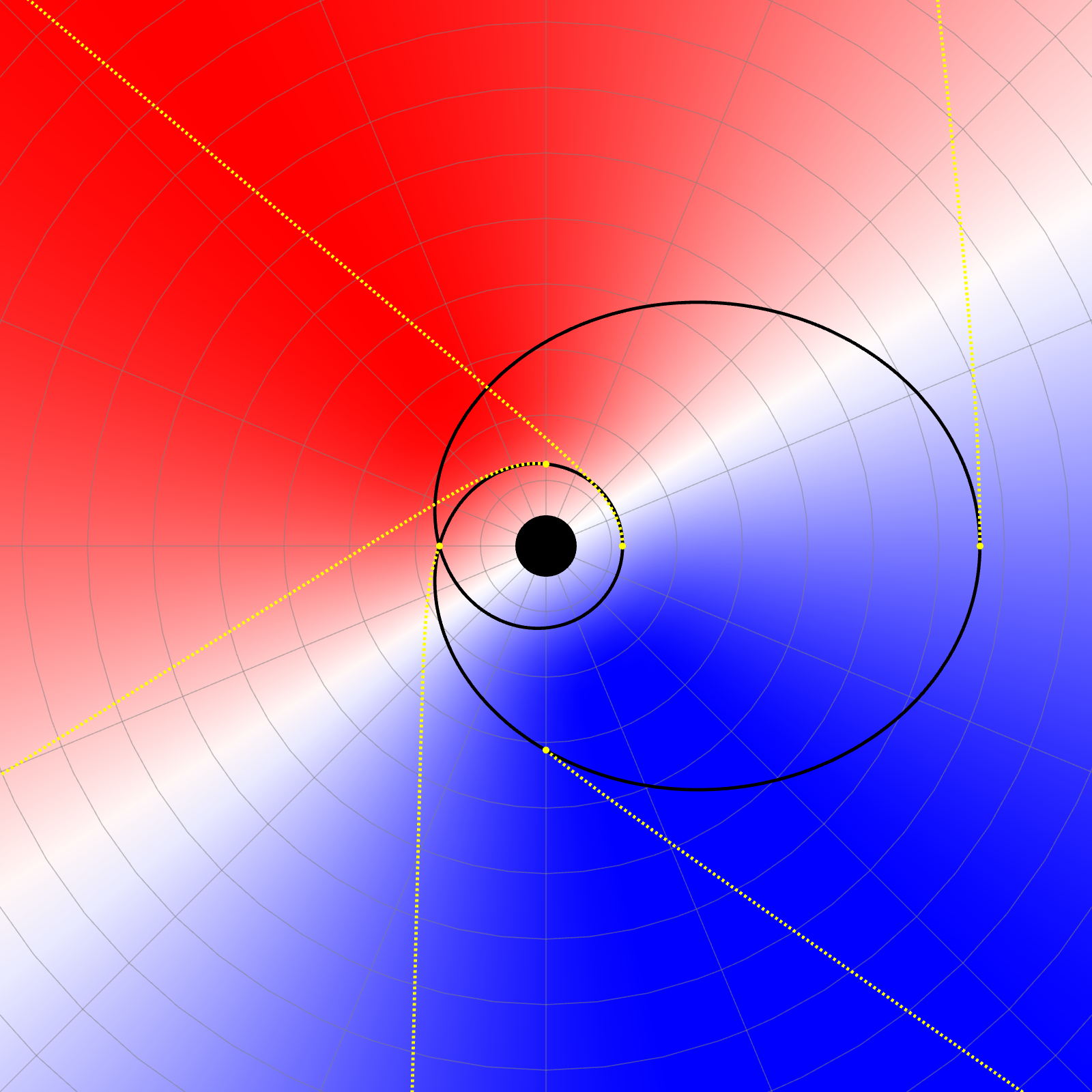}
\caption{The distribution of gravitational radiation from a 1:2 $r\phi$-resonant orbit around a Schwarzschild ($a=0$) orbit. The shading indicates the deviation from the average. Also shown are the paths of null rays emitted tangentially from the orbit.}\label{fig:orbdistrays}
\end{figure} This is illustrated in Fig. \ref{fig:orbdistrays}, in which a 1:2 resonant orbit with $e=0.7$ around a Schwarzschild black hole is depicted. The shading indicates the hot and cold spots in the gravitational radiation. Also shown are null rays starting tangentially to the orbit. One sees that the ray starting from periapsis bends approximately to the maximum of the hotspot. Around highly spinning black holes the prograde resonant orbits lie closer to the black hole and the bending effect is therefore stronger.

\begin{figure}[t]
\includegraphics[width=\columnwidth]{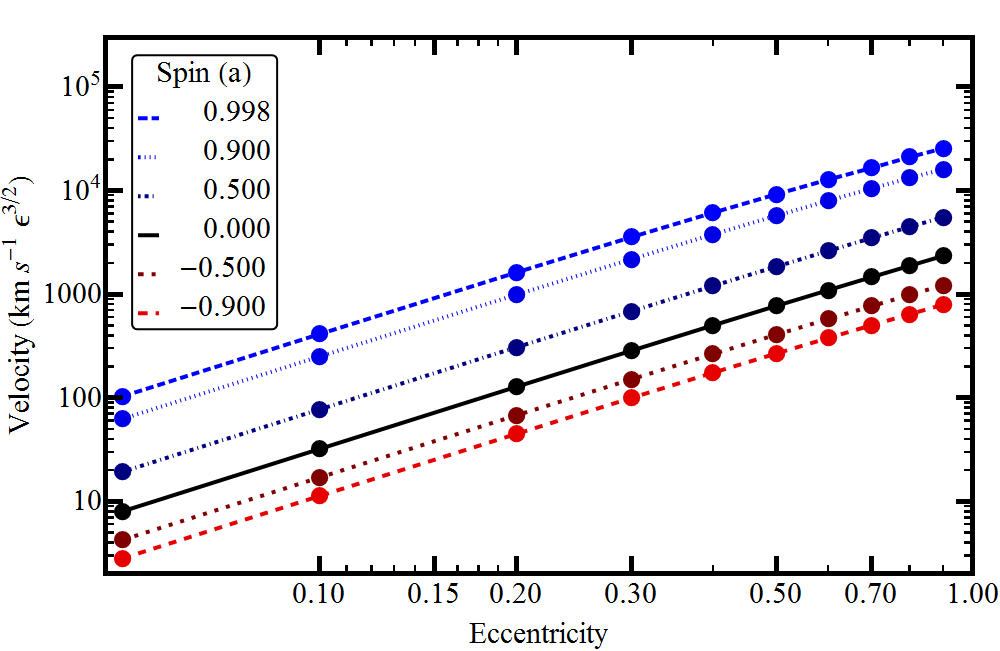}
\caption{The total magnitude of the recoil velocity accumulated by an inspiral as it evolves through a 1:2 $r\phi$-resonance, as a function of the eccentricity of the orbit and the spin of the central object.}\label{fig:kickmagplot}
\end{figure}
The total kick to the velocity of the center-of-mass of the binary system as it evolves through a resonance is shown in Fig. \ref{fig:kickmagplot}. The dependence on the eccentricity is very close to $e^2$. Only for orbits close to $e=1$ is there a slight deviation. The dependence on the spin of the central object is quite strong with $r\phi$-resonances around a spin $a=0.998$ black hole producing a kick that is ten times bigger than  $r\phi$-resonances around a Schwarzschild black hole. Not only are $r\phi$-resonances around spinning black holes more efficient (as seen in Fig. \ref{fig:radeff}), but they also produce more radiation in total. This is again a result from the  $r\phi$-resonances occurring closer to the black hole. Finally, due to the hangup effect, inspirals with aligned spin and angular momentum evolve slower than if the spin and angular momentum were unaligned. Consequently, the highly spinning inspiral spends more cycle near resonance. For high eccentricity resonances around highly spinning black holes the kick velocity can reach $30,000\,\e^{3/2}$ km/s.

\begin{figure}[t]
\includegraphics[width=\columnwidth]{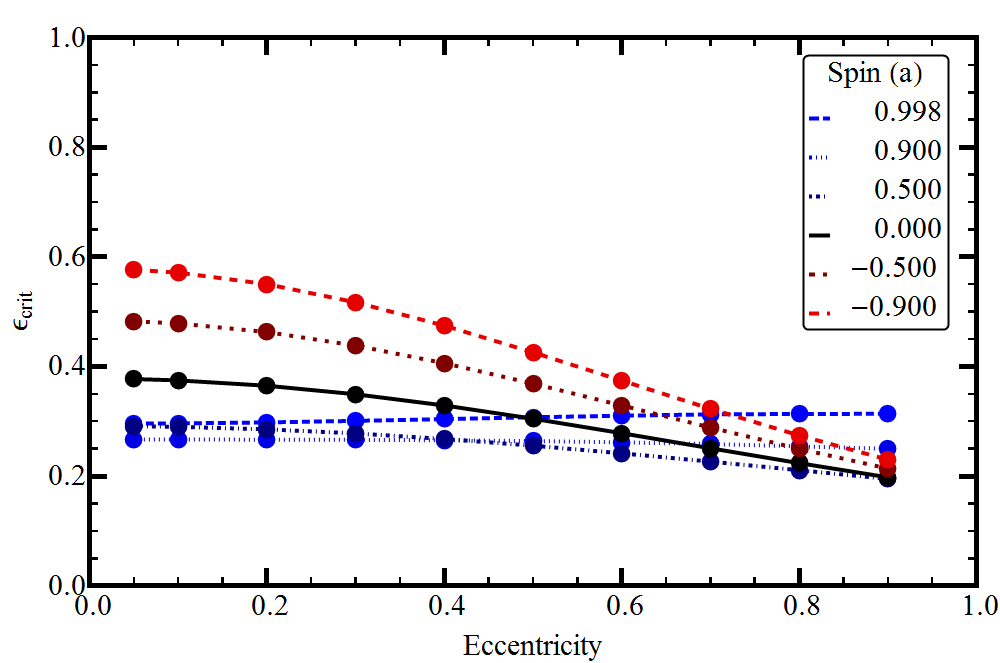}
\caption{The critical mass-ratio $\e_\mathrm{crit}$ at which evolution through an 1:2 $r\phi$-resonance can no longer be considered slow, and the analysis done in this paper breaks down.}\label{fig:epscrit}
\end{figure}

As noted at the end of section \ref{sec:linmom}, the analysis here breaks down for low mass ratios. Figure \ref{fig:epscrit} plots the value of $\e_\textrm{crit}$ as defined in eq. \ref{eq:epscrit}. The results here should thus be valid (upto order of magnitude) to about mass-ratios of 1:10. For the most extreme case this would result in a kick of almost 1,000 km/s. 

\begin{figure}[t]
\includegraphics[width=\columnwidth]{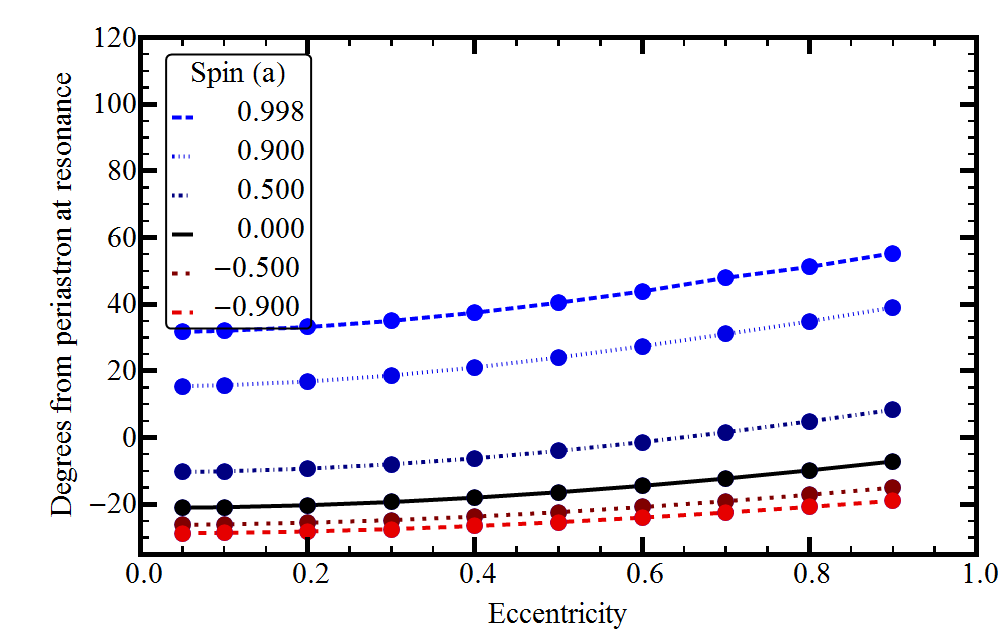}
\caption{The direction of the kicks shown in Fig. \ref{fig:kickmagplot}, measured as the angle from the periapsis as the system passes through the resonance.}\label{fig:kickangleplot}
\end{figure}

Figure \ref{fig:kickangleplot} gives the direction of the total kick. It closely mirrors Fig. \ref{fig:inbaldir}. In fact, Eq. \eqref{eq:kickvel} tells us that the direction from the total kick is simply the direction of the radiational imbalance at resonance shifted by exactly 135 degrees. That this shift is exact is a result of only one harmonic appearing due to rotational symmetry of the background. Higher order (in $\e$) corrections would also deviate from this exact shift. 
\begin{figure}[t]
\includegraphics[width=\columnwidth]{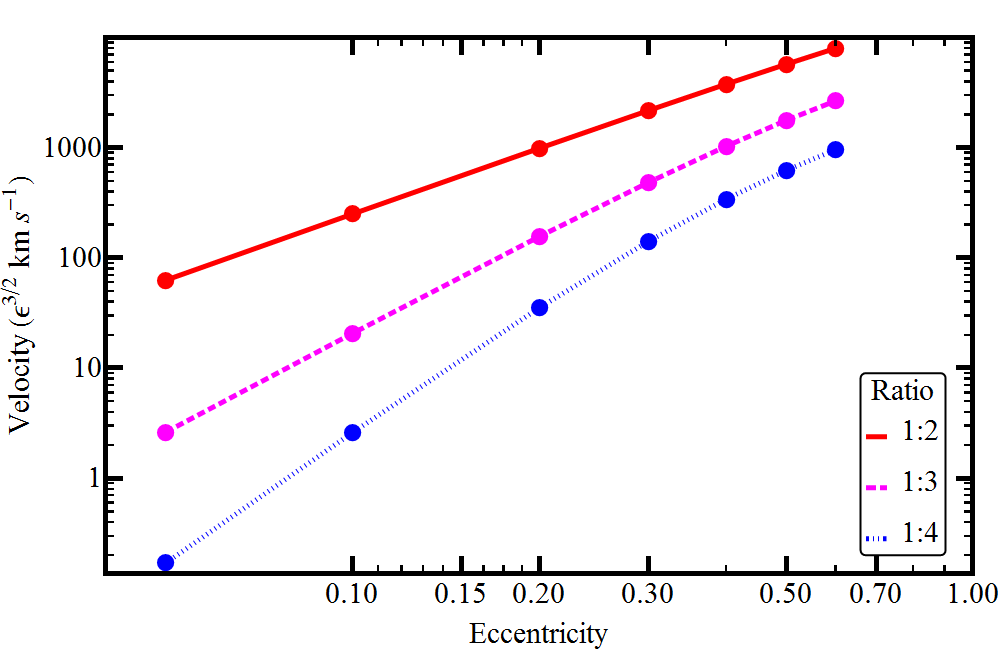}
\caption{The magnitude of the kick produced by evolution through higher order $r\phi$-resonances. As expected, the higher order resonances appear to be exponentially suppressed.}\label{fig:higherres}
\end{figure}

Thus far, we have only considered $r\phi$-resonances with a 1:2 frequency ratio. Figure \ref{fig:higherres} show the kick velocities generated by $r\phi$-resonances with different ratios around a spin $a=0.9$ central black hole. The higher order resonances are much weaker than the 1:2 resonance, despite being closer to the central black hole and producing more gravitational radiation. This is a result of the zoom-whirl behaviour of the higher order resonances; most of the gravitational radiation is emitted as the object whirls around an almost circular orbit close to the central object, while most of the asymmetry comes from the zoom that occurs once in the radial period.

The first $r\phi$-resonance encountered by an inspiral is therefore also the most potent. The effect of all subsequent resonances is likely only a fraction of the 1:2 resonance. It is also worth noting that there is no expected correlation between the phases of the different resonances encountered by an inspiral. Consequently, the sequence of kicks generated by the resonances (and the final kick from the merger) get added incoherently.

\section{Discussion and Conclusions}\label{sec:conclusions}
In this paper we have examined the effect of resonances between the radial and azimuthal periods in equatorial inspirals around a spinning central black hole. Although, due to rotational invariance one would expect very little effect on the evolution of the system, we find that the linear momentum expelled in the form of gravitational waves is enhanced significantly. In the most extreme case---i.e. an extremely eccentric inspiral around a nearly maximally spinning black hole---this can produce a kick to the velocity of the center-of-mass of the system up to $5,000\;(\e/\e_\textrm{crit})^{3/2}$ km/s. This is strong enough to dislodge an intermediate mass black hole from a host globular cluster or even in an extreme case a massive black hole from its host galaxy.

However, in reality inspirals with such extreme eccentricities and spins are almost certainly extremely rare. Nonetheless, it is conceivable that a non-negligible fraction of inspirals produces noticeable kicks through this mechanism. If it happens only once in the evolution of an intermediate mass black hole it could have significant impact on their further evolution. A more proper assessment of the astrophysical impact of this type of resonantly enhanced kick requires a more in depth analysis of the expected distribution of mass-ratios, eccentricities and spins of the relevant inspirals. This will be addressed in future work.

In this paper we examined resonant enhancement of kicks in equatorial orbits (i.e. the spin of the central object is aligned with the orbital angular momentum). The $r\phi$-resonances studied here generically occur also for inclined orbits. However, due to precession of the orbital plane, the imbalance in the gravitational wave flux gets smeared out of a larger area, weakening the kick effect. We thus expect the resonantly enhanced kick effect from $r\phi$-resonances to be maximal for equatorial orbits. This in contrast to the ``superkicks'' \cite{Campanelli:2007ew,*Campanelli:2007cga,Gonzalez:2007hi} that occur for merging highly spinning comparable mass black holes, which require the spins to lie in the orbital plane.

It is worth reiterating that the results in this paper were obtained using black hole perturbation theory by expanding to the lowest relevant order in the mass ratio $\e$. However, at the values of $\e_\mathrm{crit}$ found in Fig. \ref{fig:epscrit}, $\e$ is no longer very small, and one expects higher order corrections in $\e$ to become relevant. Nonetheless, the leading order result should still be indicative of the order of magnitude of the produced kicks.

\begin{acknowledgments}
The author would like to thank Leor Barack, Sarp Akcay and Abhay Shah for helpful discussions. He also thanks Scott Hughes and William Throwe for answering some of his questions regarding their implementation of their Teukolsky solver. This work was supported by a Rubicon grant from the Netherlands Organisation for Scientific Research (NWO).
\end{acknowledgments}

\appendix
\section{Total evolution of uncoupled constant of motion through resonance}\label{app:Xkick}

Let $X$ be a constant of motion that is uncoupled from the equations of motion (i.e. $X$ does not appear in the right hand side of any of the equations of motion for $\vec\U$ or $\vec\q$). In general, the evolution equation for $X$ can be written,
\begin{align}
\d{X}{\mt} &= \e^a F(\vec\U,\vec\q) +\bigO(\e^{a+1})\\
&= \e^a\sum_{n,k,m} F_{mkn}(\vec\U)\ee^{\ii( n\q_r+k \q_\theta+m\q_\phi)}+\bigO(\e^{a+1}),
\end{align}
with $a$ the lowest order of $\e$ at which corrections appear. Away from any resonance the oscillating terms can be moved to higher order (in $\e$) corrections by  applying a near identity transformation \cite{KC:1996,Meent:2013}
\begin{equation}\label{eq:NIAT}
\tilde{X} = X +\!\!\sum_{n,k,m}\!\!\frac{\ii\e^a  F_{mkn}(\vec\U)}{n\U_r+k\U_\theta+m\U_\phi}\ee^{\ii( n\q_r+k \q_\theta+m\q_\phi)},
\end{equation} 
yielding the adiabatic approximation
\begin{equation}\label{eq:Xadeom}
\d{\tilde{X}}{\mt} = \e^a F_{000}(\vec\U)+\bigO(\e^{a+1}).
\end{equation} 

Near a resonance $\U_\perp\equiv n_\perp\U_r+k_\perp\U_\theta+m_\perp\U_\phi$ vanishes for  some set of integers $(n_\perp,k_\perp,m_\perp)$, and consequently the corresponding oscillating terms cannot be removed by the above procedure due to the appearance of $\U_\perp$ in the denominator of the transformation. Nonetheless, the near-identity averaging transformation can still be used to remove all other (non-resonant) oscillating terms from the equation,
\begin{equation}\label{eq:Xavgeom}
\d{X}{\mt} =  \e^a F_{000}(\vec\U) +\e^a\sum_{N\neq 0} F_N(\vec\U) \ee^{\ii N\qp}+\bigO(\e^{a+1}), 
\end{equation}
where $\qp \equiv n_\perp \q_r + k_\perp \q_\theta+m_\perp \q_\phi$, $F_N \equiv F_{N n_\perp,N k_\perp,N m_\perp}$, and the integers $(n_\perp,k_\perp,m_\perp)$ have been chosen such that they have no common divisors and $n_\perp$ is positive (or when $n_\perp=0$, $k_\perp$ is positive). Also, for the sake of brevity of notation, we have dropped the tilde on the transformed $X$.
  
We are specifically interested in the case where $ F_{000}(\vec\U)$ vanishes. In the more general case, we can replace $X$ by the deviation of $X$ from the adiabatic approximation $X_{ad}$ (defined as the solution to eq. \eqref{eq:Xadeom}), $X-X_{ad}$.

We can analyze the evolution through resonance by introducing a new boundary layer time \cite{Meent:2013},
\begin{equation}
\mti = \ei\mt,
\end{equation}
with $\ei=\e^{1/2}$, and expanding $\vec\U$ and $\qp$ in $\ei$,
\begin{subequations}\label{eq:transexp}
\begin{align}
\U_i(\mt,\ei) &= \Ui_i^0+\ei \Ui^1_i(\mti) + \bigO(\ei^2),\\
\qp(\mt,\ei) &=\qpi^0(\mti) + \bigO(\ei).
\end{align} 
\end{subequations} 
Inserting these expansions in \eqref{eq:Xavgeom}, we find to lowest order in $\e$,
\begin{equation}
\d{ X}{\mti} = \e^{a-1/2}\sum_{N\neq 0} F_N(\vec\Ui^0) \ee^{\ii N\qpi^0}+\bigO(\e^{a}).
\end{equation} 
In general, $\qpi^0$ satisfies
\begin{equation}\label{eq:qpi0eom}
\dd{\qpi^0}{\mti}=G_\perp(\vec\Ui^0,\qpi^0),
\end{equation} 
and we can find the total change of $X$ incurred as the system evolves through resonance can be calculated by integrating,
\begin{equation}\label{eq:totalDX1}
\Delta{X} = \e^{a-1/2}\int_{-\infty}^{\infty} \sum_{N\neq 0} F_N(\vec\U_0) \ee^{\ii N \qpi(\mti)} \id{\mti}+\bigO(\e^{a}).
\end{equation}

If $G_\perp(\vec\Ui^0,\qpi^0)$ has a constant value $G_0$ with respect to $\qpi^0$ (as is the case for $r\phi$ and $\theta\phi$ resonances), then \eqref{eq:qpi0eom} has a simple solution,
\begin{equation}
\qpi^0(\mti) = \qp(0)+ \frac{G_0}{2}\mti^2.
\end{equation} 
With this solution for $\qpi^0$ the integrals in \eqref{eq:totalDX1} can be evaluated explicitly,
\begin{align}\label{eq:totalDX2}
\Delta{X} &=\e^{a-\frac{1}{2}} \sum_{N\neq 0}\frac{\sqrt{2\pi}F_N(\vec\U_0) }{\abs{N G_0}^{1/2}}\ee^{\ii N\qp(0)\pm\ii\frac{\pi}{4}},
\end{align}
where the $\pm$ sign is given by the sign of $N G_0$.

A crucial assumption in the above derivation is that the evolution of the system through resonance is slow compared to the orbital timescale. Without this assumption the near identity averaging transformation \eqref{eq:NIAT} breaks down, i.e. the higher order $\e$ correction terms will be of similar size as the leading order contribution. A measure for the time scale of the resonance is the dephasing time $\Lambda_{\mathrm{dep}}$, i.e. the time needed for the resonant phase $\qp$ to change $\pi/2$ from the value at resonance,
\begin{equation}
\Lambda_{\mathrm{dep}} = \sqrt{\frac{4\pi}{\abs{\d{\U_\perp}{\mt}}}}.
\end{equation}
We thus obtain that the dephasing time  $\Lambda_{\mathrm{dep}}$ compared to $\Lambda_r=2\pi/\U_r$ has to be small,
\begin{equation}
 \sqrt{\frac{4\pi}{\abs{\d{\U_\perp}{\mt}}}}\ll\frac{2\pi}{\U_r}.
\end{equation} 
This can be rephrased as a upper limit on the mass-ratio $\e$ for which this analysis is valid,
\begin{equation}
\e \ll \e_{\mathrm{crit}}\equiv\frac{\U_r^2}{\pi \abs{\avg{G_\perp}}}.
\end{equation}
\bibliographystyle{apsrev4-1}
\bibliography{journalshortnames,linkick}

\end{document}